\documentclass[a4paper,11pt,dvips,oneside]{article}
\usepackage{booktabs,graphicx,overcite,xspace,longtable}
\usepackage{graphicx,xspace}
\usepackage[dvips]{color}
\usepackage[scanall]{psfrag}

\begin{document}

\title{Some Further Results for the Stationary Points and Dynamics of Supercooled Liquids}
\author{David J.~Wales and Jonathan P.~K.~Doye \\
{\it University Chemical Laboratories, Lensfield Road,} \\
{\it Cambridge CB2 1EW, UK} }
\maketitle

\begin{abstract}
We present some new theoretical and computational results for the stationary points of bulk systems. 
First we demonstrate how the potential energy surface can be partitioned into catchment basins
associated with every stationary point using a combination of Newton-Raphson and eigenvector-following
techniques.
Numerical results are presented for a 256-atom supercell representation of a binary Lennard-Jones system.
We then derive analytical formulae for the number of stationary points as a function of both system
size and the Hessian index, using a framework based upon weakly interacting subsystems.
This analysis reveals a simple relation between the total number of stationary points, the number of local
minima, and the number of transition states connected on average to each minimum.
Finally we calculate two measures of localisation for the displacements corresponding to Hessian
eigenvectors in samples of stationary points obtained from the Newton-Raphson-based geometry optimisation
scheme. Systematic differences are found between the properties of eigenvectors corresponding to positive
and negative Hessian eigenvalues, and localised character is most pronounced for stationary points with
low values of the Hessian index.
\end{abstract}

\section{Introduction}
\label{sec:intro}

Stationary points of a potential energy surface, $V$, correspond to structures where the gradient of the
potential vanishes, while the Hessian index, $I$, is defined 
as the number of negative Hessian eigenvalues.
Stationary points with $I\ge1$ are often referred to as saddles.
A number of studies have recently focused upon the properties of saddle points for model 
supercooled liquids and glasses.\cite{AngelaniDRSS00,BroderixBCZG00,Cavagna01}
A rigorous partitioning of the potential energy surface is possible in terms of the catchment
basins of local minima [often termed `inherent structures' for bulk material\cite{StillW82,StillW84a}] 
using steepest-descent paths. 
The superposition approach to thermodynamics, where the total partition function is written
as a sum over local minima, follows naturally from this 
scheme.\cite{stillingerw82,stillingerw84,Stillinger88,stillinger95,walesdmmw00,DebenedettiS01,Wales03}
Dynamical properties can also be calculated within this framework if minimum-to-minimum rate constants
are available, using master equation,\cite{kampen81,kunz95} 
kinetic Monte Carlo,\cite{BortzKL75,Gillespie76,Gilmer80,FichthornW91} 
or discrete path sampling\cite{Wales02} techniques.
These schemes generally consider only local minima and true transition states, i.e.~stationary points
with Hessian index one (a single negative eigenvalue) in view of the 
Murrell-Laidler theorem,\cite{murrelll68}
which states that if two minima are connected by an index two saddle, then there must be a lower energy path
between them involving only true transition states.

The potential energy surface itself is independent of temperature, atomic masses and the coordinate system.
Direct analysis of $V$ itself has revealed connections between the global properties of the surface and the
interparticle forces,\cite{Wales01}
as well as a theoretical basis 
for the empirical result\cite{HoareM83,Tsai93a} that the number of different local
minima increases exponentially with size.\cite{stillingerw82,stillingerw84,stillinger99,DoyeW02}
The distribution of  minima as a function of the potential energy can also be found
from simulation data,\cite{doyew95b,Wales03} and this approach has now been used in several 
studies of bulk material.\cite{heuer97,stillinger98,Schulz98,buchnerh99,SchroterD99,StillingerSTR99,
sciortinokt99,SciortinoKT00,coluzzipv00b,DebenedettiS01,Sastry01,StarrSLSSS01}
An approximately Gaussian form is expected for sufficiently large
systems.\cite{SpeedyD95,Speedy96,HeuerB00,DoyeW02}
Distributions of transition states and barrier heights have also been reported
for bulk models.\cite{MousseauB00,HernandezW01,middletonw01,middletonw03}

Proposals to extend such ideas to consider a partition of configuration
space in terms of saddle points have been suggested in the context of
supercooled liquids and glasses.\cite{Cavagna01}
This approach has much in common with the instantaneous normal mode theory
developed by Keyes and coworkers,\cite{madank93,Keyes97,DonatiST00}
where the focus of attention is the spectrum of Hessian eigenvalues for instantaneous configurations.
Theoretical contributions based upon the premise of 
dividing configuration space into
saddles have appeared,\cite{Cavagna03,Parisi03,Grigera03,ShellDP03}
following the observation that the Hessian index of the stationary points sampled in simulations
approaches zero around the critical temperature of mode-coupling 
theory.\cite{AngelaniDRSS00,BroderixBCZG00,DiLeonardo01,AngelaniDPR01,Scala01,ShahC01,GrigeraCGP02,ShahC02,AngelaniDRSS02,
SampoliBEAR03,Angelani03b}
However, in contrast with the partitioning scheme based upon local minima, 
a well-defined procedure for using other stationary points in this way
has yet to be fully developed. In particular, the most common mapping
used is the minimisation of the square gradient, $|\nabla V|^2$,\cite{GNewton} 
but this tends not to produce true stationary points for large systems, as we
have explained in a previous contribution.\cite{DoyeW02}
In \S \ref{sec:NR} we show how an alternative approach based upon Newton-Raphson and
eigenvector-following techniques can be used to achieve the desired partitioning.

A further development of the present work is the analysis of stationary points for bulk models
in terms of independent subsystems. 
In \S \ref{sec:enumerate} we first present analytic solutions for the combinatorial problem that
determines the number of stationary points of Hessian index $I$ in a system containing $N$ atoms.
These results complement recent theoretical developments 
of Keyes and coworkers\cite{Keyes00,KeyesCK02}
based upon a random energy model approach,\cite{Derrida80,Derrida81,BryngelsonW87,onuchicwls97}
and should be useful in future studies that consider the dynamics and thermodynamics
of model systems.\cite{walesd01}
We then consider the validity of the independent subsystem approximation by examining properties of
the Hessian eigenvectors for the samples of stationary points reported in \S \ref{sec:NR}.

Since saddles beyond index one usually make no significant contribution to the dynamics of
small molecules it is important to demonstrate that
conventional dynamical approaches are not incompatible
with sampling of higher index saddles in simulations.
Just as most of the volume of a hypersphere is associated with the surface when the
dimension becomes high enough,\cite{ma85}
most of the configuration space in a large system must lie near the surface of the catchment basins for
the local minima.\cite{KurchanL96}
This observation does not invalidate the partition of configuration space into the basins of attraction
of local minima. Nevertheless it has been suggested that as the temperature increases the dynamics of a
supercooled liquid might be interpreted in terms of transitions between configuration space associated with
saddles, rather than between local minima.\cite{AngelaniDRSS00,BroderixBCZG00,
DiLeonardo01,AngelaniDPR01,Scala01} 
Such a partitioning is now technically feasible, as shown in \S \ref{sec:NR}, 
but it is not yet clear whether it offers any advantages over the simpler approach based on minima and true
transition states.

\section{A New Partitioning Scheme}
\label{sec:NR}

In this section we describe a partitioning scheme that divides the potential energy surface into regions
associated with each stationary point. Numerical results have been obtained for the same binary Lennard-Jones
system of 256 atoms (205 A and 51 B) as we considered in earlier work,\cite{DoyeW02}
and further details of the model can be found in that publication.
Qualitative changes in behaviour for various properties
have previously been reported for this system around the mode-coupling theory critical
temperature of $T_c\approx0.435$, as mentioned above.

In addition to the new searches described below,
all the molecular dynamics (MD) and geometry optimisations 
were repeated for the 256-atom system to correct a systematic problem that affected the
results in our previous report.\cite{DoyeW02} 
The present results therefore supersede those in
the latter paper, although none of the conclusions in the previous contribution
are affected.
As before, every 1000th configuration from the data
collection phase was saved and used as a starting point for the following geometry optimisations:
(1) minimisation using Nocedal's LBFGS algorithm,\cite{lbfgs}
(2) a transition state search using hybrid eigenvector-following,\cite{munrow99,walesdmmw00,kumedamw01}
(3) minimisation of $|\nabla V|^2$ using Nocedal's LBFGS algorithm.\cite{lbfgs}
The implementation of these algorithms and the convergence criteria were the same as
in previous work.\cite{DoyeW02}
The new geometry optimisations in the present work are based upon a combination of Newton-Raphson-type steps
and eigenvector-following. 
The step along eigendirection $\alpha$ in the Newton-Raphson-type search was taken 
as\cite{walesw96,walesdmmw00}
\begin{equation}
x_\alpha = {\displaystyle -2 g_\alpha \over 
\displaystyle \varepsilon^2_\alpha(1+\sqrt{1+4 g_\alpha^2 /\varepsilon^4_\alpha})},
\end{equation}
where $g_\alpha$ and $\varepsilon^2_\alpha$ are the component of the gradient and Hessian eigenvalue
corresponding to eigenvector $\alpha$, respectively. 
The sign of $\varepsilon^2$ determines whether the step in direction $\alpha$ raises or lowers the energy,
and leads to the well-known result that Newton-Raphson-type geometry optimisations can converge to stationary
points of any Hessian index.\cite{wales92,wales93d,Wales03}
In the present case this is precisely the behaviour that we require, and each Newton-Raphson-type
optimisation therefore began with a maximum of twenty such steps in combination with a dynamic
trust radius scheme for the maximum step size.\cite{Fletcher80,simonsjto83,walesw96}
Some of these initial geometry optimisations were already
converged to a root-mean-square gradient of less than
$10^{-7}\,\epsilon_{\rm AA}/\sigma_{\rm AA}$. However, for configurations obtained from the higher
temperature simulations the usual behaviour was for the number of negative Hessian eigenvalues to fall
somewhat from the initial value and then oscillate without converging. This behaviour probably arises because
the geometry optimisation does not proceed systematically uphill or downhill when the sign of an eigenvalue
changes sign. We therefore employed a fixed number 
(twenty) of these Newton-Raphson-type steps before switching to
an eigenvector-following search for a stationary point of fixed Hessian index given by the number of negative
eigenvalues at the last Newton-Raphson iteration. The step now becomes
\begin{equation}
x_\alpha = {\displaystyle \pm2 g_\alpha \over 
\displaystyle |\varepsilon^2_\alpha|(1+\sqrt{1+4 g_\alpha^2 /\varepsilon^4_\alpha})},
\end{equation}
plus for uphill and minus for downhill, independent of the sign of $\varepsilon^2_\alpha$, and all these
searches were tightly converged to stationary points of the required index.
A dynamic trust radius step scaling scheme was again employed in this phase of the optimisation.
 
The results of these calculations are collected in Tables \ref{tab:MD}, \ref{tab:diffs}
and \ref{tab:ivalues}. As in previous work the number of different minima and transition states
sampled and the fraction of negative eigenvalues
decrease markedly around $T_c$, but the system is only trapped in a single local minimum
on the simulation time scale well below this temperature.\cite{HernandezW01,middletonw01,DoyeW02,middletonw03}
Most minimisations of $|\nabla V|^2$ converge to non-stationary-points (NSP's) rather than true
stationary points (SP's) of $V$, especially at higher temperature.

Statistics for 
the mean potential energy difference and displacement in configuration 
space between the starting point and the
final geometry after each optimisation are collected in Table \ref{tab:diffs}.
While the displacements are practically the same for each class of geometry optimisation the energy
differences are always ordered 
$\Delta V_{\rm min} > \Delta V_{\rm ts} > \Delta V_{\rm G2} > \Delta V_{\rm NR}$.
By this measure the Newton-Raphson-type scheme has clearly succeeded in finding `closer' stationary points
to the original configurations, although the displacement statistics
do not discriminate between them.
This scheme  also provides a true partitioning of the configuration space 
into regions associated with all the
stationary points, although the precise boundaries will depend upon details of how the algorithms are
implemented, such as the maximum step size allowed.\cite{wales92,wales93d}
In contrast, if steepest-descent paths are used to divide the space into catchment basins for
local minima then the mapping is in principle unambiguous. This result follows because
steepest-descent is defined in terms of a first-order differential equation,
for which a uniqueness theorem applies.\cite{Wales03}

\section{Stationary Points for Independent Subsystems}
\label{sec:enumerate}

We now extend previous results\cite{stillingerw82,stillingerw84,stillinger99,DoyeW02} 
for the number of local minima and transition states
to stationary points of any Hessian index.
This analysis is based upon the assumption that a sufficiently large system of
$m N$ atoms can be divided into $m$ effectively independent subsystems of $N$ atoms each.
Writing the number of stationary points of index $I$ in an $N$-atom system as
$n_{\rm sp}(N,I)$ it follows that
\begin{equation}
n_{\rm sp}(m N,0)=n_{\rm sp}(N,0)^m.
\label{eq:min_mN}
\end{equation}
The solution of this equation is
\begin{equation}
n_{\rm sp}(N,0)=\exp(\alpha N),
\end{equation}
where $\alpha$ is a constant.

A similar argument can be given for the number of transition states.\cite{DoyeW02}
Assuming that the rearrangements associated with the transition states can 
be localised to one subcell, the whole $mN$-atom system will be at a 
transition state when one of the subsystems is at a transition state 
and the rest are at a minimum, so
\begin{equation}
\label{eq:ts_mN}
n_{\rm sp}(m N,1)=m\,n_{\rm sp}(N,1) n_{\rm min}(N)^{m-1},
\end{equation}
with solution
\begin{equation}
n_{\rm sp}(N,1)=aN \exp(\alpha N),
\end{equation}
where $a$ is a another constant.

For index two saddles we have
\begin{eqnarray}
\label{eq:sp2_mN}
n_{\rm sp}(m N,2)=m\,n_{\rm sp}(N,2)\, n_{\rm sp}(N,0)^{m-1} + 
{m(m-1)\over 2}\,n_{\rm sp}(N,1)^2\, n_{\rm sp}(N,0)^{m-2}, 
\end{eqnarray}
where the two terms correspond to the two ways that an index two saddle
for the $mN$-atom system can be generated.
More generally 
\begin{equation}
\label{eq:spI_mN}
n_{\rm sp}(m N,J)=m! \sum_{\{n_0,n_1,\cdots,n_J\}} 
\prod_{I=0}^J \frac{n_{\rm sp}(N,I)^{n_I}}{n_I!}, 
\end{equation}
where $n_I$ is the number of the $m$ subsystems that are at a stationary 
point of index $I$, and the sum is over the combinations of $n_I$ values that 
satisfy $\sum_I n_I=m$ and $\sum_I I\, n_I=J$. The $m!/\prod_{I=0}^J n_I!$
factor in the above sum is a multinomial coefficient, and the sum
itself corresponds to the terms in the multinomial series expansion of
\begin{equation}
\label{eq:multi}
\left[n_{\rm sp}(N,0)+n_{\rm sp}(N,1)+ \cdots +n_{\rm sp}(N,I_{\rm max})\right]^m
\end{equation}
with the appropriate value of $J$, where $I_{\rm max}$ is the maximum value 
of $I$ for the $N$-atom system.
This observation enables us to derive a general expression for $n_{\rm sp}(N,I)$
by writing the multinomial as
\begin{equation}
\left[x^0 n_{\rm sp}(N,0)+x^1 n_{\rm sp}(N,1)+ x^2 n_{\rm sp}(N,2) + \cdots +
x^{I_{\rm max}} n_{\rm sp}(N,I_{\rm max})\right]^m.
\end{equation}
The terms contributing to $n_{\rm sp}(mN,J)$ correspond to the coefficient of $x^J$, so
\begin{equation}
n_{\rm sp}(mN,J) = \left.\frac{1}{J!} \frac{d^J}{dx^J} 
\left[\sum_{I=0}^J x^I n_{\rm sp}(N,I) \right]^m\right|_{x=0}.
\end{equation}
We can now prove that the general solution to these equations is
\begin{equation}
\label{eq:gensol}
n_{\rm sp}(N,I)= {aN \choose I} \exp(\alpha N),
\end{equation}
where the binomial coefficient 
\begin{equation}
{aN \choose I}={aN!\over I! (aN-I)} 
\ \ \hbox{or}\ \ {\Gamma(aN+1)\over \Gamma(I+1) \Gamma(aN-I+1)!} 
\end{equation}
for integer and non-integer values of $a$, respectively.
Substituting for $n_{\rm sp}(N,I)$ from equation (\ref{eq:gensol}) gives
\begin{eqnarray}
n_{\rm sp}(mN,J) &=& \left.\frac{1}{J!} \frac{d^J}{dx^J} 
\left[\sum_{I=0}^J x^I {aN \choose I} \exp(\alpha N)  \right]^m\right|_{x=0} \nonumber \\
&=&  \left.\frac{1}{J!} \frac{d^J}{dx^J} (1+x)^{amN}  \right|_{x=0}  \exp(\alpha mN) \nonumber \\
&=&  \frac{J!}{J!} {amN \choose J} \exp(\alpha mN) = {amN \choose J} \exp(\alpha mN),
\end{eqnarray}
which is consistent with the supposition of equation (\ref{eq:gensol}).

From equation (\ref{eq:multi}) it is clear that the total number of 
stationary points of any index must obey
\begin{equation}
n^{\rm tot}_{\rm sp}(mN)= n^{\rm tot}_{\rm sp}(N)^m.
\end{equation}
Therefore, $n^{\rm tot}_{\rm sp}(N)$, like $n_{\rm sp}(N,0)$, is a simple 
exponential:
\begin{equation}
n^{\rm tot}_{\rm sp}(N)= \exp(\gamma N),
\end{equation}
where $\gamma$ is a constant.

$n^{\rm tot}_{\rm sp}(N)$ can also be obtained from $n_{\rm sp}(N,I)$ by summing over $I$:
\begin{eqnarray}
n^{\rm tot}_{\rm sp}(N)
                    &=& \exp(\alpha N) \sum_I {aN \choose I} 
                    = 2^{aN} \exp(\alpha N) = \exp\left[(\alpha +a \ln 2) N\right]. 
\end{eqnarray}
Our result for $n_{\rm sp}(N,I)$ 
therefore implies a relationship between the the total number of stationary points, the scaling exponent
for the number of minima and the number of transition states connected
on average to each minimum, which is $n_{\rm sp}(N,1)/n_{\rm sp}(N,0)=aN$.
These parameters are connected by the equation
\begin{equation}
\gamma=\alpha +a \ln 2,
\end{equation}
which we anticipate may play an important role in future energy landscape analysis of bulk systems.
For example, the ratio between the number of transition states and the number of local minima must
scale appropriately for thermodynamic and dynamic properties to exhibit proper extensive or
intensive behaviour.\cite{Keyes00,walesd01,KeyesCK02,ShellDP03}

If $n_{\rm sp}(N,I)$ has the form given in equation (\ref{eq:gensol}) then
the probability of choosing a stationary point of index $I$ from all the stationary points of a system
containing $N$ atoms is
\begin{equation}
P(N,I) = {aN \choose I} 2^{-aN}.
\end{equation}
This distribution can be closely approximated by the Gaussian form
\begin{equation}
\label{eq:Gauss}
P(N,I) \approx \frac{e^{-(I-\mu)^2/2\sigma^2}}{\sqrt{2\pi\sigma^2}},
\end{equation}
with $\mu=Na/2$ and $\sigma^2=Na/4$. Although this approximation does not give the correct scaling
behaviour for $P(N,1)/P(N,0)$ it is quite accurate in most other respects.
The difference between the tails of the two distributions occurs because the 
Gaussian tends to zero at $I\rightarrow\pm\infty$, whereas
the binomial distribution vanishes at $I$=$-1$ and $aN+1$,
so that $I_{\rm max}$=$aN$. The latter limit provides a physical constraint
on $a$, namely, $a\le d$, where $d$ is the dimension of the system. 

The above results complete our 
previous analysis, and are in good agreement with the Gaussian distributions
found empirically for small clusters.\cite{DoyeW02} For the clusters the 
maximum of the distribution occurs at $I\approx N-2$, and therefore 
$a\approx 2$.
Whether this result holds for bulk systems with periodic boundary conditions
is not yet clear. The value $a=2$ may reflect the open boundary conditions 
for the clusters, where there must always be $N-1$ stable modes 
or the cluster will dissociate.\cite{DoyeW02} 

Our results also complement the empirical observation that the 
average potential energy of stationary points with index $I$ appears 
to scale linearly with $I$. 
A theoretical justification for the latter observation has been obtained by
Shell {\it et al.\/}~within a similar framework that considers independent
subsystems.\cite{ShellDP03}
These authors employ a maximum term approximation to find the most 
probable saddle index, and equation (\ref{eq:Gauss}) indicates that 
this should be a good approximation for large enough systems.
In particular, they predict that the density of saddle points with fractional
index, $i=I/dN$, should scale exponentially with size with a 
coefficient that depends on $i$, i.e.\ as 
$\exp\left[N\theta(i)\right]$. Using our expression for $n_{\rm sp}(N,I)$ 
it is easy to show that 
\begin{equation}
\theta(i)=\alpha+{1\over N} \ln {aN \choose idN},
\end{equation}
which varies between $\alpha$ and $\alpha+a\ln 2$ (in the limit of large $N$).
Our derivations also complement results based on the Morse rules,\cite{morse34}
which can provide upper and lower bounds for the number of stationary points of a given
index.\cite{mezey81a,mezey87}

\section{Stationary Points and Dynamics}
\label{sec:local}

The above analysis is based on the assumption of independent, or weakly interacting, subsystems.
Previous work on finite size effects for model glass formers provides support for this picture,
including the result that `a system of $N=130$ particles behaves basically as two non-interacting
systems of half the size'.\cite{buchnerh99,DoliwaH03b}
In this idealised limit the procedure to calculate dynamical properties, such as the diffusion constant, $D$,
is straightforward. For example, suppose that each subsystem consists of
minima that are each connected to two neighbouring minima via two transition states. 
If the rate constant corresponding to each transition state is the same, and equal to $k$, then the expected
waiting time between transitions is $1/2k$. 
If the minima are assumed to be arranged linearly in space, and each connected pair is separated by the
same distance $d$, then the mean squared displacement after $n$ steps
is simply $nd^2$ for uncorrelated transitions.
The average time for $n$
steps to occur is $n/2k$, and so the one-dimensional diffusion constant is simply $D=kd^2$. 
For a weakly interacting system of $m$ such subsystems the dimension of the configuration space is multiplied
by $m$. The expected waiting time in any one of these 
higher-dimensional minima is now $1/2mk$, because there are now $2m$
transition states connected to each minimum. However, the same diffusion constant is obtained as for a single
subsystem, because the displacement must be averaged over $m$ times as many atoms.

Now suppose that we try to find the diffusion constant using a kinetic Monte Carlo 
approach\cite{BortzKL75,Gillespie76,Gilmer80,FichthornW91}
and calculations of minima and transition states.\cite{HernandezW01}
In this case a statistical rate theory may be used to estimate the transition state $k$ using the properties
of the stationary points or pathways.
For the single subsystem described above the average waiting time per transition is then $1/2k$,
and the correct diffusion constant will be obtained so long as
both transition states are located for each minimum visited.
Now consider the system containing $m$ weakly interacting copies of this subsystem. 
In this case it is necessary to
locate $2m$ transition states per minimum, and the waiting time becomes $1/2mk$, so that much smaller time
increments occur in the resulting KMC simulation. However, the correct diffusion constant should still be
obtained after averaging displacements over all the subsystems.

\begin{figure}[h]
\psfrag{0}[cr][cr]{0}
\psfrag{0.05}[cr][cr]{0.05}
\psfrag{0.1}[cr][cr]{0.10}
\psfrag{0.15}[cr][cr]{0.15}
\psfrag{0.2}[cr][cr]{0.20}
\psfrag{0.3}[cr][cr]{0.30}
\psfrag{I}[cr][cr]{$I$}
\psfrag{0b}[tc][tc]{0}
\psfrag{0.5}[tc][tc]{0.5}
\psfrag{1b}[tc][tc]{1}
\psfrag{1.5}[tc][tc]{1.5}
\psfrag{2}[tc][tc]{2}
\psfrag{temp}[tc][tc]{$kT/\epsilon$}
\psfrag{1}[cr][cr]{1}
\psfrag{0.01}[cr][cr]{}
\psfrag{0.0001}[cr][cr]{}
\psfrag{1e-05}[cr][cr]{$10^{-5}$}
\psfrag{1e-06}[cr][cr]{}
\psfrag{1e-08}[cr][cr]{}
\psfrag{1e-10}[cr][cr]{$10^{-10}$}
\psfrag{5}[tc][tc]{}
\psfrag{10}[tc][tc]{10}
\psfrag{15}[tc][tc]{}
\psfrag{20}[tc][tc]{20}
\psfrag{25}[tc][tc]{}
\psfrag{30}[tc][tc]{30}
\psfrag{35}[tc][tc]{}
\psfrag{40}[tc][tc]{40}
\psfrag{45}[tc][tc]{}
\psfrag{50}[tc][tc]{50}
\psfrag{1overT}[tc][tc]{$\epsilon/kT$}
\centerline{\includegraphics[width=14 cm]{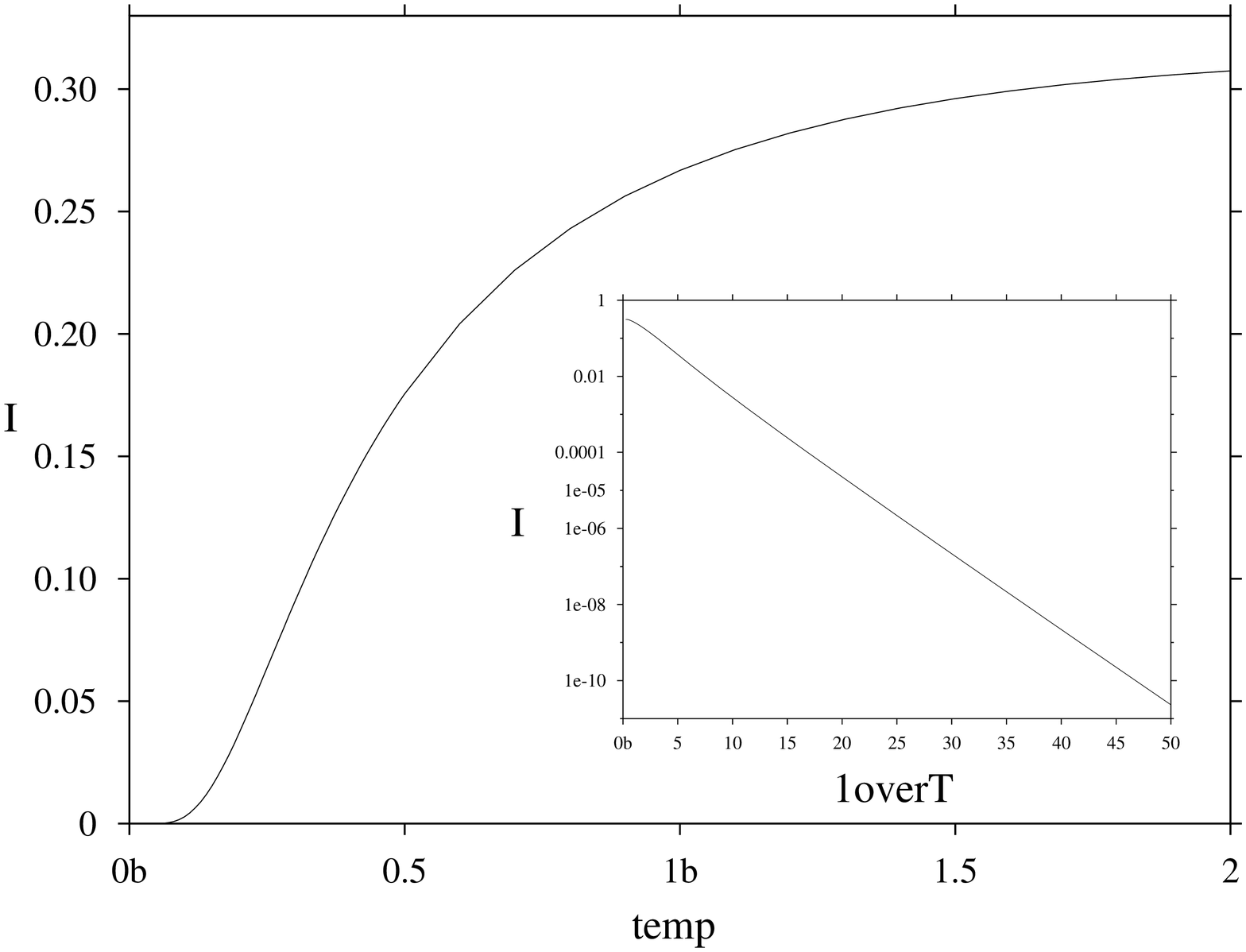}}
\ \vfill
\caption{\label{fig:IofT}
}
\end{figure}

It is clear from the above discussion that KMC simulations may become 
much less efficient as the system size
increases if all the atoms play an active role in diffusion. At higher temperatures efficiency is not really
the point, since standard MD calculations provide a 
straightforward route to $D$. Rather, the KMC approach is of
interest because it provides a coarse-grained picture of the relevant processes. Sampling problems do indeed
appear to be an issue for KMC simulations of glasses, as we will report elsewhere.\cite{MiddletonW03b}
Nevertheless, the correct diffusion constant can in principle be obtained from such calculations by
considering only local minima and true transition states, as long as the 
transitions are Markovian. This result is not in contradiction with the
observation that progressively higher index saddles may be sampled as the temperature increases. For example,
the mean Hessian index of instantaneous configurations or stationary points obtained by the geometry
optimisation procedure of \S \ref{sec:NR} can easily be calculated for one-dimensional 
double-well (Figure \ref{fig:IofT}) or periodic 
potentials.\cite{AngelaniDRSS02,Angelani03c} 
The average index for $m$ copies of such non-interacting systems would simply be $m$ times larger.
Hence the stationary points sampled by the Newton-Raphson procedure of \S \ref{sec:NR} 
would include progressively higher index saddles as a function of temperature, up to a limiting
value determined by $m$. 
It may also be significant that the index falls to zero only in the zero 
temperature limit, although linear extrapolation of data from the range $0.2<kT/\epsilon<0.5$ 
would suggest a non-zero value (Figure \ref{fig:IofT}). 
Interestingly, as noted previously from simulations by Doliwa and Heuer\cite{Doliwa03a}
and from simple models by Berthier and Garrahan,\cite{Berthier03} if the same
data is represented on an Arrhenius plot a straight line results, arguing
against a sharp transition in the properties of saddles near to the
mode-coupling temperature.

It is sometimes stated that the inherent structure approach
to the dynamics of supercooled liquids is only
relevant in the temperature range where there is a separation 
of time scales for vibrational motion and 
transitions between minima.\cite{Cavagna01,SchroderSDG00,DennyRB03}
However, the above discussion on the system size dependence of the 
residence times for local minima indicates
that this separation depends on system size as well as the temperature. 
Goldstein, in his landmark paper on the application 
of the potential energy landscape to the study of supercooled 
liquids\cite{Goldstein69}, actually 
argued for the separation of two {\em atomic} time scales. 
The first time scale corresponds to 
localised vibrations of an atom around its mean position, 
and the second to the time taken for a transition that 
involves a significant displacement of this atom. 
Goldstein realized that, whilst a specified atom is vibrating,
many transitions could potentially occur between minima 
that involve rearrangements localised elsewhere in the system.
In practice, the inherent structure approach to dynamics will be most useful
when the times between interminimum transitions are long compared 
to the time scale for vibrational motion. For this reason 
it has been argued that the system size needs to be 
small enough for this time scale separation to be
present.\cite{BuchnerH00,KeyesC01,DennyRB03,Berthier03b} 

The previous discussion also 
shows that analysis of dynamics using a configuration space partitioned into local
minima and considering only true transition states is not incompatible with the system `sampling' 
higher index saddles. However, all the arguments are based on the notion of independent subsystems, and we 
therefore examine how well this picture might apply to the stationary points obtained for the binary
Lennard-Jones system discussed in \S \ref{sec:NR} using the Newton-Raphson-type procedure.
Two distinct indices have been calculated for each of these points, as described below.

\begin{figure}[h]
\psfrag{0.60}[cr][cr]{0.60}
\psfrag{0.55}[cr][cr]{0.55}
\psfrag{0.50}[cr][cr]{0.50}
\psfrag{0.45}[cr][cr]{0.45}
\psfrag{0.40}[cr][cr]{0.40}
\psfrag{0.35}[cr][cr]{0.35}
\psfrag{0.30}[cr][cr]{0.30}
\psfrag{0.6}[cr][cr]{0.6}
\psfrag{0.5}[cr][cr]{0.5}
\psfrag{0.4}[cr][cr]{0.4}
\psfrag{0.3}[cr][cr]{0.3}
\psfrag{0.2}[cr][cr]{0.2}
\psfrag{0.1}[cr][cr]{0.1}
\psfrag{0.0}[cr][cr]{0.0}
\psfrag{0}[tc][tc]{0}
\psfrag{5}[tc][tc]{5}
\psfrag{10}[tc][tc]{10}
\psfrag{15}[tc][tc]{15}
\psfrag{20}[tc][tc]{20}
\psfrag{25}[tc][tc]{25}
\psfrag{30}[tc][tc]{30}
\psfrag{local}[cr][cl]{$L$}
\psfrag{Ntilde}[cr][cl]{$\widetilde N$}
\psfrag{200}[cr][cr]{200}
\psfrag{160}[cr][cr]{160}
\psfrag{120}[cr][cr]{120}
\psfrag{80}[cr][cr]{80}
\psfrag{40}[cr][cr]{40}
\psfrag{0}[cr][cr]{0}
\psfrag{Hessian index/i}[tc][tc]{Hessian index}
\psfrag{plus}[tc][tc]{plus}
\psfrag{minus}[tc][tc]{minus}
\centerline{\includegraphics[width=12 cm]{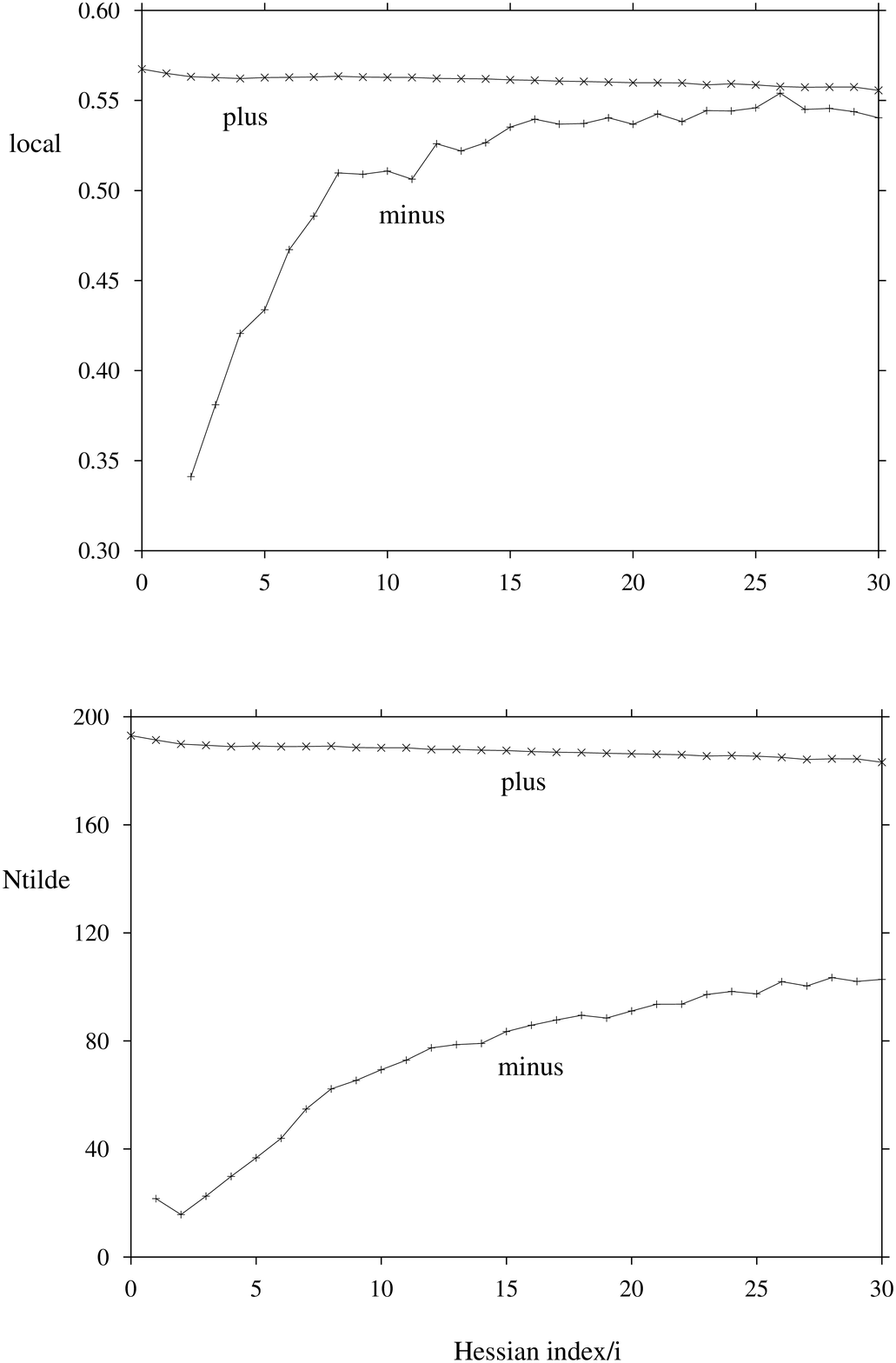}}
\ \vfill
\caption{\label{fig:local}
}
\end{figure}

The first `localisation' index, $L$, was defined as
\begin{equation}
L_{\alpha\beta} = \sum_{\gamma =1}^{3N} \left|c_\gamma^\alpha c_\gamma^\beta\right|,
\end{equation}
where $c_\gamma^\alpha$ is component $\gamma$ of normalised Hessian eigenvector $\alpha$. 
For every stationary point with Hessian index two or higher we calculated the average value
of $L$ over the $I(I-1)/2$ pairs of eigenvectors with negative eigenvalues, $L_{\rm minus}$.
The statistics for stationary points of the same index were found to be
very similar for samples obtained from the MD runs at different temperatures, 
and so averages over all the 
runs are presented in Figure \ref{fig:local}. For eigenvectors corresponding to motion in different
regions of space (or different atoms)
we would expect $L=0$, while for motion localised on the same atoms $L$ should
approach unity. 
The corresponding averages over all pairs of eigenvectors corresponding to positive eigenvalues,
$L_{\rm plus}$, were also calculated for comparison.
Figure \ref{fig:local} reveals that $L_{\rm plus}$ is around 0.55
practically independent of the index. In contrast, $L_{\rm minus}$ is systematically smaller, particularly
when the index is less than about ten.

We have also calculated 
\begin{equation}
\widetilde N_\alpha =  \left( \sum_\gamma \left(c^\alpha_\gamma\right)^2 \right)^2 \Big/
\sum_\gamma  \left(c^\alpha_\gamma\right)^4 ,
\label{eq:coop}
\end{equation}
for all eigenvectors with non-zero Hessian eigenvalues.
$\widetilde N$ is proportional to the participation ratio,\cite{NagelGR84}
and is expected to vary between one for localised modes to about $N$ for delocalised modes.
Figure \ref{fig:local} shows the results for modes with positive and negative eigenvalues,
$\widetilde N_{\rm plus}$ and $\widetilde N_{\rm minus}$, separately for comparison,
and again we display averages over stationary points from the different MD runs.
Here the difference between $\widetilde N_{\rm plus}$ and $\widetilde N_{\rm minus}$ is even more
marked than for $L_{\rm plus}$ and $L_{\rm minus}$.

The statistics for both $L$ and $\widetilde N$ both indicate that the characteristic displacements associated
with Hessian eigenvectors that have negative eigenvalues are more localised and spatially independent than
for eigenvectors associated with positive eigenvalues. For both measures it is the stationary points with the
fewest negative eigenvalues for which this character is most pronounced.
These results appear to agree very well with the analysis of Shell {\it et al.\/},
who conclude that low-index saddles may often be described in terms of combinations of 
transition states of the subsystems.\cite{ShellDP03}

\section{Conclusions}
\label{sec:conclude}

We have previously pointed out the need for rate constants and related dynamical properties, such as the
diffusion constant, to scale correctly with system size.\cite{DoyeW02}
The total energy and its fluctuations are extensive quantities, while the
barrier heights between local minima and true transition states are intensive.
If `activated' processes are defined by comparing such extensive and intensive quantities
then one would be forced to conclude that no 
`activated' processes exist at any infinitesimal temperature in 
a bulk system.\cite{AngelaniDRSS00,SampoliBEAR03}
We have previously argued that such a definition is inappropriate,\cite{DoyeW02} 
since rate constants for well-defined
geometrical rearrangements are intensive quantities, as are the expressions used to calculate them in standard
unimolecular rate theory.\cite{baerh96}
In this sense all transitions between the catchment basins of local minima are `activated', since a potential
energy barrier is involved, although a more useful definition should probably consider the magnitude of
$kT$ or the available energy per degree of freedom.

Our previous results,\cite{HernandezW01,middletonw01,DoyeW02,middletonw03}
now supported by independent calculations,\cite{DoliwaH03}
indicate that the relevant potential energy barriers for diffusion are generally not small 
compared to $kT$ in the supercooled region. In fact,
statistical rate theories have been successfully applied to diffusion in solids for 
nearly fifty years, and
the standard approach is based on transition state theory for true transition states.\cite{Vineyard57}
At first sight, these observations might appear to be incompatible with the notion that the system samples
mostly higher index saddles above some temperature threshold. 
However, the analysis of \S \ref{sec:local} indicates that if the potential energy surface is
partitioned into catchment basins of local minima in the usual way then the dynamics can indeed be treated by
considering the lowest barriers between them, which are those mediated by true transition
states.\cite{murrelll68}
Of course, the usual caveats apply to such an analysis, namely that the transitions are assumed to be
Markovian, and a statistical theory is usually
employed to calculate the required rate constants, often involving a
harmonic approximation.

We have shown how an alternative partitioning scheme based upon Newton-Raphson and eigenvector-following
geometry optimisation can successfully divide the potential energy surface into catchment basins associated
with all the stationary points. Thermodynamic\cite{ShellDP03}
and dynamic\cite{Keyes00,KeyesCK02,ShellDP03} schemes might be constructed on this basis along similar lines
to the methods used for the conventional partitioning in terms of local minima.
However, the latter division is simpler, and the theoretical tools that use it are comparatively well
developed.\cite{Wales03}

For weakly interacting subsystems we have now solved the combinatorial problem
that defines the number of stationary points of any given index. 
This analysis also reveals a simple relation between the parameters that determine
the total number of stationary points, the number of local minima, and the number of transition
states connected on average to each minimum.
We have further investigated the displacements corresponding to the Hessian eigenvectors of all the
stationary points located using the Newton-Raphson-based scheme of \S \ref{sec:NR} to see
whether they are compatible with the above analysis.
The results indicate that eigenvectors corresponding to negative eigenvalues 
involve displacements where fewer atoms participate than for eigenvectors
with positive eigenvalues, and that the displacements corresponding to different eigenvectors are
more independent. Both these characteristics are most pronounced for stationary points with
low values of the Hessian index, in agreement with previous work in which such points are
considered as combinations of true transition states for subsystems.\cite{ShellDP03}

\section*{Acknowledgements}
JPKD is grateful to the Royal Society for the award of a University Research Fellowship.


\newcommand\aciee{Angew. Chem. Int. Ed. Engl.\xspace}
\newcommand\ac{Acta. Crystallogr.\xspace}
\newcommand\acp{Adv. Chem. Phys.\xspace}
\newcommand\acr{Acc. Chem. Res.\xspace}
\newcommand\ajp{Am. J. Phys.\xspace}
\newcommand\ap{Ann. Physik\xspace}
\newcommand\arpc{Ann. Rev. Phys. Chem.\xspace}
\newcommand\bbpc{Ber. Bunsenges. Phys. Chem.\xspace}
\newcommand\bc{Biochemistry\xspace}
\newcommand\bmk{Biometrika\xspace}
\newcommand\bp{Biopolymers\xspace}
\newcommand\cccc{Coll. Czech. Chem. Comm.\xspace}
\newcommand\cop{Comm. Phys.\xspace}
\newcommand\cp{Chem. Phys.\xspace}
\newcommand\cpc{Comp. Phys. Comm.\xspace}
\newcommand\cpl{Chem. Phys. Lett.\xspace}
\newcommand\crev{Chem. Rev.\xspace}
\newcommand\ea{Electrochim. Acta\xspace}
\newcommand\el{Europhys. Lett.\xspace}
\newcommand\epjd{Eur. Phys. J. D\xspace}
\newcommand\fd{Faraday Discuss.\xspace}
\newcommand\ic{Inorg. Chem.\xspace}
\newcommand\ijmpc{Int. J. Mod. Phys. C\xspace}
\newcommand\ijqc{Int. J. Quant. Chem.\xspace}
\newcommand\jcis{J. Colloid Interface Sci.\xspace}
\newcommand\jcsft{J. Chem. Soc., Faraday Trans.\xspace}
\newcommand\jacers{J. Am. Ceram. Soc.\xspace}
\newcommand\jacs{J. Am. Chem. Soc.\xspace}
\newcommand\jas{J. Atmos. Sci.\xspace}
\newcommand\jcc{J. Comp. Chem.\xspace}
\newcommand\jchp{J. Chim. Phys.\xspace}
\newcommand\jcp{J. Chem. Phys.\xspace}
\newcommand\jce{J. Chem. Ed.\xspace}
\newcommand\jcscc{J. Chem. Soc., Chem. Commun.\xspace}
\newcommand\jetp{J. Exp. Theor. Phys. (Russia)\xspace}
\newcommand\jmb{J. Mol. Biol.\xspace}
\newcommand\jmsp{J. Mol. Spec.\xspace}
\newcommand\jmst{J. Mol. Struct.\xspace}
\newcommand\jncs{J. Non-Cryst. Solids\xspace}
\newcommand\jpa{J. Phys. A\xspace}
\newcommand\jpb{J. Phys. B\xspace}
\newcommand\jpc{J. Phys. Chem.\xspace}
\newcommand\jpca{J. Phys. Chem. A\xspace}
\newcommand\jpcb{J. Phys. Chem. B\xspace}
\newcommand\jpcm{J. Phys.: Condens. Matt.\xspace}
\newcommand\jpcs{J. Phys. Chem. Solids.\xspace}
\newcommand\jpsj{J. Phys. Soc. Jpn.\xspace}
\newcommand\jsp{J. Stat. Phys.\xspace}
\newcommand\jvsta{J. Vac. Sci. Technol. A\xspace}
\newcommand\mg{Math. Gazette\xspace}
\newcommand\molphys{Mol. Phys.\xspace}
\newcommand\molp{Mol. Phys.\xspace}
\newcommand\nat{Nature\xspace}
\newcommand\nsb{Nature Struct. Biol.\xspace}
\newcommand\pac{Pure. Appl. Chem.\xspace}
\newcommand\pd{Physica D\xspace}
\newcommand\phys{Physics\xspace}
\newcommand\pma{Philos. Mag. A\xspace}
\newcommand\pmagb{Philos. Mag. B\xspace}
\newcommand\pnas{Proc. Natl. Acad. Sci. USA\xspace}
\newcommand\pnasu{Proc. Natl. Acad. Sci. USA\xspace}
\newcommand\ppmsj{Proc. Phys.-Math. Soc. Japan\xspace}
\newcommand\pr{Phys. Rev.\xspace}
\newcommand\prep{Phys. Reports\xspace}
\newcommand\pra{Phys. Rev. A\xspace}
\newcommand\prb{Phys. Rev. B\xspace}
\newcommand\prbcm{Phys. Rev. B\xspace}
\newcommand\prc{Phys. Rev. C\xspace}
\newcommand\prd{Phys. Rev. D\xspace}
\newcommand\pre{Phys. Rev. E\xspace}
\newcommand\prl{Phys. Rev. Lett.\xspace}
\newcommand\prsa{Proc. R. Soc. A\xspace}
\newcommand\psfg{Proteins: Struct., Func. Gen.\xspace}
\newcommand\ptps{Prog. Theor. Phys. Supp.\xspace}
\newcommand\rmp{Rev. Mod. Phys.\xspace}
\newcommand\sci{Science\xspace}
\newcommand\sa{Sci. Amer.\xspace}
\newcommand\ssci{Surf. Sci.\xspace}
\newcommand\tca{Theor. Chim. Acta\xspace}
\newcommand\zpb{Z. Phys. B.\xspace}
\newcommand\zpc{Z. Phys. Chem.\xspace}
\newcommand\zpd{Z. Phys. D\xspace}
\newcommand\zpdamc{Z. Phys. D\xspace}
\newcommand\phm{Philos. Mag.\xspace}

\bibliographystyle{thesis}

\begin{thebibliography}{10}
\bibitem{GNewton}
 The only exception to the use of the $|\nabla V|^2$ mapping, as far as we
  know, is the use of a `Newton' method by Grigera {\it et
  al.}\cite{GrigeraCGP02}, but only a brief outline of the method was given.
\end{thebibliography}

\renewcommand{\tabcolsep}{1.25 mm}
\begin{table*}[h]
\begin{center}
\caption{\label{tab:MD}
Mean total energy, $E$, potential energy, V, kinetic energy, KE, and kinetic 
equipartition temperature, $T$, for seven
MD  simulations of a BLJ system with number density 1.2\,$\sigma_{\rm AA}^{-3}$.\cite{Wales03}
A cutoff of $2.5\,\sigma_{AA}$ was employed in these calculations together with a shifting scheme that
ensured continuity of the energy and gradient.\cite{DoyeW02}
The $\pm$ values represent one standard deviation. $\#{\rm min}$, $\#{\rm ts}$,
$\#{\rm G2}$ and $\#{\rm NR}$ are the number of distinct minima,
transition states, stationary points of  $|\nabla V|^2$
and stationary points of $V$, found for $10^3$ searches,
using L-BFGS minimisation,\cite{Nocedal80,lbfgs}
hybrid eigenvector-following transition state searching,\cite{munrow99}
minimisation of $|\nabla V|^2$ using the L-BFGS algorithm,
and Newton-Raphson-type steps, respectively (excluding permutational isomers).
\%SP and \%NSP are the percentage of quenches on the $|\nabla V|^2$ surface that converged to
stationary points and non-stationary points of $V$, respectively (out of $10^3$ total).\cite{DoyeW02} }
\bigskip
\bigskip
\centerline{
\begin{tabular}{ccllccccccc}
\hline
 run & $E$  & \qquad $V$ & \ \ KE & $T$ & $\#{\rm min}$ & $\#{\rm ts}$ & $\#{\rm G2}$ & $\#{\rm NR}$ & \%SP & \%NSP \\
\hline
1 & $-1648.904\pm0.003$ & $-1723\pm3$ & \ $75\pm3$  & $0.196\pm0.007$ &   1 &  188 & 85 &87&  17.8 & 82.2 \\
2 & $-1599.986\pm0.005$ & $-1700\pm4$ &  $100\pm4$  & $0.262\pm0.005$ &  14 &  198 & 419 &379&  12.6 & 87.4 \\
3 & $-1499.959\pm0.008$ & $-1652\pm6$ &  $152\pm6$  & $0.399\pm0.015$ & 592 &  874 & 998 &995&  2.3 & 97.7 \\
4 & $-1399.959\pm0.012$ & $-1602\pm7$ &  $202\pm7$  & $0.530\pm0.020$ & 555 &  892 & 1000 &1000&  1.2 & 98.8\\
5 & $-1299.946\pm0.015$ & $-1536\pm9$ &  $236\pm9$  & $0.619\pm0.024$ & 998 & 1000 & 1000 &1000&  1.5 & 98.5\\
6 & $-1199.951\pm0.020$ & $-1481\pm11$ & $281\pm11$ & $0.737\pm0.028$ & 1000 & 1000 & 1000 &1000& 2.7 & 97.3\\
7 & $-1099.952\pm0.025$ & $-1427\pm12$ & $327\pm12$ & $0.859\pm0.032$ & 1000 & 1000 & 1000 &1000& 2.3 & 97.7\\
\hline
\end{tabular} }
\end{center}
\end{table*}

\renewcommand{\tabcolsep}{2.00 mm}
\begin{table*}[h]
\begin{center}
\caption{\label{tab:diffs}
Mean potential energy differences, $\Delta V$, and displacements, $\Delta D$, 
between the starting point and the converged geometry
after searching for minima (min), transition states (ts), 
minimising $|\nabla V|^2$ (G2), and performing Newton-Raphson-type (NR) optimisation.\cite{Wales03}
The $\pm$ values represent one standard deviation. 
}
\bigskip
\bigskip
\centerline{
\begin{tabular}{crrrrrrrr}
\hline
           run & $\Delta V_{\rm min}$ & $\Delta D_{\rm min}$\quad & $\Delta V_{\rm ts}$ & $\Delta D_{\rm ts}$ \
               & $\Delta V_{\rm G2}$  & $\Delta D_{\rm G2}$  
               & $\Delta V_{\rm NR}$  & $\Delta D_{\rm NR}$ \\
\hline
1 &  $49\pm2$ & $32\pm0$ &  
     $44\pm3$ & $33\pm1$ & 
     $48\pm2$ & $32\pm0$ & 
     $47\pm5$ & $32\pm1$ \\
2 &  $98\pm4$ & $33\pm1$ & 
     $96\pm5$ & $34\pm1$ &
     $96\pm4$ & $34\pm1$ & 
     $91\pm12$ & $34\pm2$ \\
3 &  $147\pm5$ & $37\pm1$ & 
     $146\pm6$ & $37\pm1$ &
     $142\pm6$ & $37\pm1$ & 
     $121\pm21$ & $38\pm2$ \\
4 &  $197\pm8$ & $56\pm1$ &
     $195\pm8$ & $56\pm1$ &
     $187\pm9$ & $56\pm1$ & 
     $159\pm22$ & $57\pm1$ \\
5 &  $235\pm9$ & $133\pm9$ &
     $232\pm9$ & $133\pm9$ &
     $205\pm10$ & $133\pm9$ & 
     $170\pm18$ & $133\pm8$\\
6 &  $283\pm9$ & $206\pm11$ &
     $280\pm11$ & $206\pm11$ &
     $234\pm12$ & $206\pm11$ & 
     $192\pm19$ & $206\pm11$\\
7 &  $333\pm13$ & $293\pm14$ &
     $329\pm13$ & $293\pm14$ &
     $263\pm14$ & $293\pm14$ & 
     $216\pm21$ & $293\pm14$\\
\hline
\end{tabular} }
\end{center}
\end{table*}
\newpage

\begin{table*}[h]
\begin{center}
\caption{\label{tab:ivalues}
$i_{\rm SP}$ and $i_{\rm NSP}$ are the fractions of negative Hessian eigenvalues found after
minimising $|\nabla V|^2$, split into stationary points and non-stationary points of $V$, 
respectively.\cite{DoyeW02}
$i_{\rm NR}$ is the corresponding fraction for Newton-Raphson-type searches.\cite{Wales03}
The $\pm$ values represent one standard deviation. 
}
\bigskip
\bigskip
\begin{tabular}{crrr}
\hline
           run & $i_{\rm NR}\times10^3$ & $i_{\rm NSP}\times10^3$ & $i_{\rm SP}\times10^3$ \\
\hline
1 &  $0.4\pm0.7$ & $0.4\pm0.7$ & $0.0\pm0.0$ \\
2 &  $1.4\pm1.1$ & $1.0\pm1.0$ & $0.6\pm0.8$ \\
3 &  $4.1\pm2.1$ & $3.6\pm1.8$ & $3.2\pm1.8$ \\
4 &  $4.7\pm2.1$ & $3.5\pm1.9$ & $3.4\pm1.5$ \\
5 &  $13.6\pm3.8$ & $10.9\pm3.2$ & $16.4\pm3.7$\\
6 &  $21.7\pm4.3$ & $17.7\pm3.8$ & $19.5\pm4.6$\\
7 &  $29.1\pm4.8$ & $23.9\pm4.4$ & $25.8\pm3.4$\\
\hline
\end{tabular}
\end{center}
\end{table*}
\clearpage

\section*{Figure Captions}

\begin{enumerate}
\item 
Mean Hessian index as a function of temperature (canonical ensemble) 
in reduced units of $kT/\epsilon$ for 
the double well potential $8\epsilon(2x^4-x^2)$, 
which has a barrier height of $\epsilon$.
The Hessian index calculated for instantaneous configurations and 
after the Newton-Raphson procedure of \S \ref{sec:NR} give 
indistinguishable results. 
The inset shows the same data presented in an Arrhenius plot. 
\item
Variation of $L$ and $\widetilde N$, parameters designed to provide insight into the localisation of
displacements along
Hessian eigenvectors, as a function of the Hessian index. Both parameters were averaged over all the 
relevant stationary points obtained by Newton-Raphson-type optimisation from configurations sampled 
in the seven MD runs described in Table \ref{tab:MD}.
For $L$ separate 
averages were calculated
over all pairs of eigenvectors corresponding to negative eigenvalues 
and all pairs corresponding to positive eigenvalues within each stationary point.
For $\widetilde N$ separate averages were 
compiled for Hessian eigenvectors corresponding to positive and
negative eigenvalues, as indicated. 
\end{enumerate}

\end{document}